# Catastrophic rupture of lunar rocks: Implications for lunar rock size–frequency distributions

Ottaviano Rüsch, Rachael M. Marshal, Wajiha Iqbal, Jan Hendrik Pasckert, Carolyn H. van der Bogert, Markus Patzek

Institut für Planetologie, Westfälische Wilhelms Universität Münster, Münster, Germany

**Abstract**

Like many airless planetary surfaces, the surface of the Moon is scattered by populations of blocks and smaller boulders. These features decrease in abundance with increasing exposure time due to comminution by impact bombardment and produce regolith. Here we model the evolution of block size–frequency distributions by updating the model of Hörz et al. (1975) with new input functions: the size–frequency distributions of cm–scale meteoroids observed over the last few tens of years and a rock impact shattering function. The impact shattering function is calibrated using measurements of a lunar block size–frequency distribution of known age. We find that cumulative block size–frequency distributions change with time from a power–law for young populations (<~50 Myr) to an exponential distribution for older populations. The new destruction rates are within the uncertainty of the original model, although, for sizes >5 cm, two times faster than the original best estimate. The faster rates are broadly consistent with observations reported by other studies. Since the input functions are known for small rock sizes, the rock abundance can be determined theoretically at sizes below the current image spatial resolution (0.5 m). Surface exposure age of block fields can be estimated together with the initial block abundance from the measurement of block size–frequency distributions.

# 1. Introduction

Blocks (1-100 m) and smaller boulders (0.1-1 m) (Bruno and Ruban, 2017) are ubiquitous on planetary surfaces as a result of impact cratering. The common approach to study these features is the measurement of their size–frequency distribution (SFD). Impact ejecta block SFD have been measured extensively on the Moon since the Surveyor era (e.g., Shoemaker and Morris, 1970), using orbital imaging data (e.g., Cintala and McBride, 1995; Bart and Melosh, 2010; De Rosa et al., 2012; Krishna and Kumar, 2016; Pajola et al., 2019; Watkins et al., 2019) and up to recent time by Chinese lander missions (e.g., Di et al., 2016; Li et al., 2017; Li et al., 2018; and Li and Wu, 2018; Wu et al., 2018; Wu et al., 2021). SFD measurement of impact ejecta blocks have been performed on many airless bodies, such as Ida (Lee et al., 1996), Phobos (Thomas et al., 2000), Eros (e.g., Thomas et al., 2001; Dombard et al., 2010; Michikami and Hagermann, 2021), Itokawa (e.g., Saito et al., 2006; Michikami et al., 2008; Mazrouei et al., 2014; DeSouza et al., 2015), Lutetia (Küppers et al., 2011), Toutatis (Jiang et al., 2015), Vesta (Schröder et al., 2021a), Ceres (Schulzeck et al., 2018; Schröder et al., 2021b), Ryugu (Michikami et al., 2019; Michikami and Hagermann, 2021; Sugimoto et al., 2021), Bennu (Della Giustina et al., 2019; Burke et al., 2021), comet 67P (e.g., Pajola et al., 2015; 2016), and Enceladus (Pajola et al. 2021). Power–laws are often fitted to these measurements and cumulative power–index steeper than -2 are measured, consistent with highly fragmented material (e.g., Hartmann, 1969; Dohnanyi et al., 1971; Michel et al., 2001; Jutzi et al., 2010). Extrapolation of the SFD measurement to small sizes have been attempted using different mathematical description of the SFD shapes, i.e., power–law (e.g., Shoemaker and Morris, 1970; Li et al., 2017, Bandfield et al., 2011, Watkins et al., 2019, Krishna and Kumar, 2016), exponential (Shoemaker and Morris, 1970; Golombek and Rapp, 1997; Di et al., 2016 and Li and Wu, 2018), and with a Weibull distribution (e.g., Schroeder et al. 2021, 2020, Pajola et al., 2016). These extrapolations have been performed to compare and validate SFD measurement with thermal observations sensitive to cm and m scale blocks (e.g., Bandfield et al., 2011), and for landing site hazard analysis (e.g., Golombek and Rapp, 1997; Golombek et al., 2003; Golombek et al., 2008; Wu et al., 2018; Ruesch et al., 2021). Additionally, the relationship between boulder size and source crater has been investigated (Bart and Melosh, 2007; Jia et al., 2019).

Despite abundant SFD measurements, the understanding of these distributions is vague. Attempts to relate a block population to its exposure age have considered the population as a whole without distinction of its size distribution (Basilevsky et al., 2013; Ghent et al., 2014; Basilevsky et al., 2015; Li et al., 2018; Watkins et al., 2019; Wei et al., 2020; Ruesch et al., 2020; Bickel et al., 2020). Among these studies, the work of Basilevsky et al. (2013) and Ghent et al. (2014) have revealed how the measured destruction rate, for all size combined, is found to be higher than predicted theoretically in the study of Hörz et al. (1975). This poor understanding is in contrast with the relatively well–known erosive processes at the lunar surface, namely shattering (e.g., Hörz et al., 1975; Hörz, 1977; Cintala and Hörz, 1992; Hörz et al., 2020; Ruesch et al., 2020) and abrasion (e.g., Shoemaker et al., 1970; Gault et al., 1972; Hörz et al., 1974; Cintala and Hörz, 1992; Rüsch and Wöhler, 2022) by impact bombardment. Thus, a natural question arises: How does the SFD of a block population on the lunar surface changes with time, in particular at small sizes? This study addresses this question by demonstrating that the evolution of block SFD can be modeled with sufficient precision (in terms of size distribution and absolute time) to allow meaningful comparison with measured block abundances.

## 2. Methods
### 2.1 Overview

The model is based on an improvement of the Monte Carlo study of Hörz et al. (1975) and exploits the advances over the last 46 years in the understanding of impact shattering and meteoroid flux. Briefly, the model of Hörz et al. (1975) simulates a surface composed of isolated blocks of the same size, formed all at the same time, and subject to bombardment by meteoroids. It tracks the energy imparted to each block by multiple meteoroids and calculate when a block accumulates sufficient energy for its complete destruction. The number of hits by a meteoroid of given size is a function of the meteoroid impact frequency, block size and exposure time. The model developed here (hereafter the updated model) follows the exact same procedure and considers i) new block shattering functions, ii) better estimates of the actual meteoroid flux, meteoroid size–frequency distribution, and meteoroid velocity, and iii) the role of secondary fragments. Thermal stresses due to diurnal temperature variations have been evoked to contribute to block erosion (e.g., Molaro and Byrne, 2012; 2017) although their relative importance is likely subordinate to meteoroid impacts (Hörz et al., 2020; Ruesch et al., 2020). Because the effects of this process are not known in details, it is currently not possible to incorporate them in the model. The reconstruction of an entire block SFD is achieved by running the model for different block sizes. In the following sections only the updates to the Hörz et al., (1975) model are presented. The reader is referred to Hörz et al. (1975) for the details of the original model. To compare the model with observations, we take advantage of the high spatial resolution images of block fields acquired by the Lunar Reconnaissance Orbiter

Narrow Angle Camera (LROC/NAC) (Robinson et al., 2010) and knowledge of the lunar chronology (Drozd et al., 1977; Stöffler and Ryder, 2001; Stöffler et al., 2006)

**2.2 Shattering energy functions**

The energy required to shatter a block is usually described with the shattering threshold, defined as the specific energy ($Q^*_s$) required to shatter a target such that the mass of the largest fragment is half the mass of the target ($M_{lf} = 0.5 M_t$) (e.g., Gault and Wedekind, 1969). It has been long known that this shattering threshold is size dependent (e.g., Gault and Wedekind, 1969; Gault et al., 1972). In addition to composition, factors such as porosity and strength are obviously key in controlling the specific energy (e.g., Jutzi et al., 2010). The shattering threshold is also a function of the damage accumulated by multiple mild impact events before actual shattering (Housen, 2009). The determination of this value for representative lunar rocks ($Q^*_{s\_moon}$) with experiments is extremely challenging due to the different rheology and petrology of lunar rocks with respect to terrestrial samples and the wide range of rock types on the Moon. A second challenge proper to this study is the unknown energy (expressed as $Q/Q^*_s$) that blocks need to experience in order to become undetectable (apparently erased) in images. To overcome such challenges, we first consider shattering functions that describe the closest analogs to lunar rocks. We then scale these shattering functions with respect to actual lunar block SFD measurements. In this way, we can take into account the unknown parameters proper to lunar rocks and the unknown detectability threshold $Q/Q^*_s$. In other words, the calibration consists in finding the appropriate scale of the shattering threshold while keeping its size dependency as reported in the literature. This size dependency is not set as a free parameter because its measure has been already reported for close analogs to lunar rocks (basalt and granite).

Two functions relating the specific energy to the target size have been proposed in the literature that are relevant for this study. Relative to the function used in the original model (Gault et al., 1972), these two functions are more strongly correlated to the size (more negatively steep in an energy versus size diagram) and require less energy per unit of mass. Benz and Asphaug (1999) calibrated their collision model using impact experiments on basaltic material of Nakamura and Fujiwara (1991) (Benz and Asphaug, 1994; Benz and Asphaug, 1999). Housen and Holsapple (1999) determined a function based on experiments of granite targets of size between 1 and 30 cm. The two functions differ most notably by the different exponent of the flaw–size distribution inside the targets (Holsapple et al., 2002). The disruption threshold is equivalent to the shattering threshold in the strength regime, and can be determined as follow (Benz and Asphaug, 1999), termed BA99:

$$Q_D^* = Q_0 \left(\frac{R_{pb}}{1cm}\right)^a + B\rho \left(\frac{R_{pb}}{1cm}\right)^b \qquad \text{eq. (1)}$$

where $R_{pb}$ is the radius of the block, and $\rho$ the density of the block (in g/cm3). $Q_0$, B, a and b are parameters that have been determined for different target properties and impact speeds (Benz and Asphaug, 1999). As in Ruesch et al. (2020), for a strong non–porous target block (basalt) we use $Q_0$=2.9e7 (erg/g), B=1.5 (erg cm$^3$/g$^2$), a=-0.35 and b=1.29 (see Jutzi et al., 2010), estimated for an impact at 5 km/s. Parameters for impacts at higher velocities, representing conditions at the lunar surface, are lacking. However, the disruption threshold is not expected to vary by more than a factor of 2–3 due to differences in velocity for the small target size in consideration (Bottke et al., 2005). The specific energy for a given target size developed by Housen and Holsapple (1999), termed HH99, is:

$$Q_S^* \propto 5 \times 10^7 D_{target}^{-0.667} \qquad \text{eq. (2)}$$

The proportionality constant in eq. (2) is valid for cgs units with $D_{target}$ expressed in cm. The $Q^*_s$ values in eq. (1) and eq. (2) are multiplied by the mass of the block to be compared to the energy delivered by impacts. Each simulated block is allowed to accumulate the energy delivered by each impact over time. When a block accumulated sufficient energy, either through one "overkill" impact or multiple mild impacts, the block is shattered or, if sufficient energy is available, catastrophically shattered, (e.g., Fujiwara et al., 1977). As previously mentioned, it is not known at which accumulated energy (Q) above ~$Q^*_s$ a block appears erased in orbital images. It is however expected that $M_{lf}/M_t$, i.e., mass of the largest fragmented normalized to the mass of the target, decreases linearly with increasing $Q/Q^*_s$ (Fujiwara et al., 1977; Benz and Asphaug, 1999; Leinhard and Stewart, 2009). For an accumulated energy of $Q/Q^*_s$=1 the mass of the largest fragment is half the mass of the initial block: in this case a block is still identifiable (Figure 2a). Undetectable ("erased") blocks have likely accumulated energy above two to several times $Q^*_s$. Because this threshold value ($Q/Q^*_s$) is not known, it is a parameter that needs to be determined through a calibration. Letting the threshold $Q/Q^*_s$ be a free parameter has also the advantage to compensate for differences in tensile strength and porosity (e.g., Jutzi et al., 2010) between the rock material used in laboratory simulations and the actual properties of lunar blocks, as well as differences in impact velocities. As a drawback, the determination of the detectability threshold $Q/Q^*_s$ through a calibration does not allow to determine precisely the actual $Q^*_{s\_moon}$ of lunar rocks that could have been compared to $Q^*_s$ of theoretical models and laboratory experiments reported in the literature. Since $Q^*_{s\_moon}$ of lunar rocks cannot be known precisely, also the ratio $M_{lf}/M_t$ above which blocks appear erased in images cannot be determined precisely.

**2.3 Secondary fragments**

Hörz et al. (1975) emphasized how their model did not simulate new "daughter" blocks, i.e., fragments of a parent shattered block that are sufficiently large to be recognizable and counted during SFD measurements. Observations have shown that before accumulating energy above the detectability threshold, blocks appear most of the time intact and only a small fraction develops intermediate families (daughter blocks) (Ruesch et al., 2020). Examples of intermediate stages are shown in Figure 1a,b,c. Therefore, SFD measurement consists in identifying and counting blocks that are mainly single, intact, "parent" blocks. It is to be emphasized that the intact state is only apparent because by accumulating impact energy, blocks gradually weaken internally and loose spalls, while morphologically still appearing in one piece. The single piece appearance might often be a remanent core that lost some of the mass during large–scale damage events (sub–catastrophic events) (Housen, 2009) (Figure 1c). Importantly, clusters of daughter fragments represent a minor fraction of an entire population (Ruesch et al., 2020). Therefore, there is a rapid transition in time between single apparently "intact" blocks and highly shattered (undetectable) blocks. In a simplified model, there is no need to track the energy of intermediate daughter fragments as long as they do not constitute a major fraction of a population. There is another key observation that justifies this simplification. The largest daughter fragments cluster around the position of the parent block (Figure 1 and Ruesch et al., 2020). This is in contrast to the experiment of Housen (2009) that show that even severe–damage events accelerate away fragments. This apparently opposing result is likely due to the experiments performed for targets without ground: for the lunar case, the largest fragments (core) are accelerated toward the ground and might sink without lateral displacement. We performed a simple experiment demonstrating that shattered rocks sitting on soil do not move (Figure 2). Therefore, it is viable to identify these instances (examples shown in Figure 1a,b,c) during SFD measurements and to consider them non "shattered" and non "erased" block in order to allow a meaningful comparison with a SFD modeled without secondary fragments. The fate for the daughter fragments of small size (<<width of parent block) is different (Figure 1d). Upon block sub–catastrophic and catastrophic shattering, these fragments are ejected (e.g., Nakamura and Fujiwara, 1991), land far from the parent block (e.g., Durda et al., 2011) and are likely to be undistinguishable from the primary block population (see distal fragments in Figure 1d). As we will demonstrate, because their size is much smaller than that of the parent block, their lifetime is correspondingly much shorter. In fact, if a period of exposure is sufficiently long for a parent block of diameter D to be catastrophically shattered, then half of that period is sufficient for shattering (and erasure) of blocks of diameter <<D. Therefore, the contribution of the small daughter fragments to the SFD evolution of a block field is negligible.

**2.4 Projectile size–frequency distributions**

To determine the kinetic energy of each impact onto a block, the mass and velocity of the impactor needs to be assessed. The mass of projectiles that play a role for the erosion of a block is determined by the block mass itself. In terms of sizes, for a block of diameter $D_b$ and disruption threshold $Q^*_s$, the diameter of the projectile is (Durda et al., 1998):

$$D_p = \left(\frac{2Q^*_s}{V_{imp}^2}\frac{\rho_t}{\rho_p}\right)^{1/3} D_b \quad \text{eq. (3)}$$

with $V_{imp}$ the impact velocity, $\rho_t$ the density of the block and $\rho_p$ the density of the projectile. Assuming that the projectile density is that of ordinary chondrites, the ratio of densities is taken as unity (e.g., Consolmagno et al., 2008). The partitioning of the impact energy into shattering energy is included in the definition of the specific energy $Q^*_s$ (Durda et al., 1998). As in Hörz et al. (1975), impact contributing as little as 10% the total energy required for shattering are considered for each block. Block sizes from 1 m to 60 m are considered, consistent with the sizes measured with orbital data. This means that projectiles imparting energy as small as 0.05x$Q^*_s$ (~4 mm) and projectiles as large as ~2 m are simulated. The contribution by larger impactors is minor as discussed in Hörz et al. (1975).

The size range of interest (0.004–2 m) is not well constrained and for the 1–~5 cm range there are no direct measurements (e.g., Drolshagen et al., 2017; Suggs et al., 2014). As shown in this work, however, the projectile SFD has a key effect in the block SFD. In the original model of Hörz et al., (1975), a projectile energy distribution was derived using the frequency of lunar microcraters. Here, three projectile SFD approximations are tested (Figure 3). Together, the three SFDs cover the scattering of observational data points of actual meteoroids (Pokorny and Brown et al., 2016; Drolshagen et al., 2017). The first is the distribution of Brown et al. (2002), termed B02, with a characteristic cumulative power–law index of -2.7 and originally defined down to a diameter of 5 cm. The distribution is extrapolated here down to smaller diameters. Recent impact crater measurement on Bennu are reported to be consistent with such extrapolation (Ballouz et al., 2020). It is the shallowest distribution tested here, even if other observations of present–day impactors indicate an even shallower distribution in the range from ~10 cm to 3-4 cm with a power law index of ~1.4 (Halliday et al., 1996; Advellidou et al., 2021). For the second SFD, termed G85–B02, the distribution of projectiles smaller than 2 cm is calculated with a steeper power–law index of -4, based on the distribution of Grün et al. (1985). This steeper distribution is based on measurement of meteoroids sizes smaller than 1 cm (Grün et al., 1985). This second distribution was originally presented as possible approximation in the study of Drolshagen et al. (2017). The third projectile SFD is the interpolation termed "1g" proposed by Drolshagen et al. (2017), termed D17 here, that uses as fix reference points the Brown et al. (2002) model at the largest size (1 m) and the Grün et al., (1985) at the smallest size (1 cm). This distribution is the simplest interpolation for the size range without measurements (1–~5 cm) that most closely follow a power law (Drolshagen et al., 2017). The slope of this third SFD is the highest and the same as in the distribution of Oberst and Nakamura (1989). The latter distribution is not considered here because it was determined for sizes larger than few m.

In the simulation, the size of each impactor is determined randomly from the distribution of possible sizes in the three SFD cases. The size distributions are binned with a constant bin size in logarithmic scale such that $D_{i+1}=D_i\sqrt{2}$, D being the diameter, starting at 0.7 m. The flux at the surface of the Moon as defined in Oberst et al. (2012) is used and is assumed spatially uniform. Latitudinal and longitudinal variation of the flux and impact velocity exist (e.g., Le Feuvre and Wieczorek, 2011; Pokorny et al., 2019; Robertson et al., 2021). The strongest variation concerns the flux between polar and equatorial regions, which ratio is estimated to be 1.2 (Robertson et al., 2021). Since we are interested in the block SFD evolution on the Moon at the global scale, we do not take into account these variations. The latitude dependency of the impact angle (e.g., Robertson et al., 2021) does not play a role because facets of blocks are oriented randomly. The projectile velocity is randomly sampled from the distribution of possible velocities for the Moon as defined in Marchi et al. (2009) and is considered independent of the meteoroid size. An additional source of impacts is provided by lunar secondaries with a power–law index of -4 and a mean impact velocity of 0.5 km/s (e.g, Costello et al., 2018; Gault et al., 1974). The flux as defined in Costello et al. (2018) was used for their modeling.

**2.5 Micrometeoroid abrasion effect**

The recent study of the abrasion rates as a function of block size of Rüsch and Wöhler (2022) indicates that the abrasion affects the topography of a block from top to bottom. As a consequence, the block diameter (measured in images) remains constant as the block height decreases. Changes in block diameter occurs only when its debris apron (fillet) onlap the block.

Therefore, abrasion effects can be neglected during block SFD modeling as long as the blocks under study are not covered by their fillets.

**2.6 Model calibration**

The updated model is based on six variables: the exposure time, the initial power law distribution (parameters *a* and *b*), the impact shattering function, the meteoroid SFD, and the detectability threshold. The first variable is usually unknown and bears the most interest. The second and third variables (*a* and *b*) can be determined through a fit to a measured block SFD. The remaining three variables are assumed to be constant across the entire lunar surface and need to be determined only once through a calibration. Before presenting the details of the calibration, it is useful to determine which block SFD shapes should be expected on the Moon.

The following review of the copious literature on SFD measurements indicates that there is evidence for two types of SFD shapes: a power–law and a convex up (exponential) shape. SFD measurements on young surfaces (<~50 Myr) determined that a power–law distribution well approximate blocks counts (Bandfield et al., 2011; Krishna and Kumar, 2016; Watkins et al., 2019; Pajola et al., 2019). To characterize the SFD of blocks without being biased by resolution effects, images acquired during the descent and in situ have been used (e.g, Li and Wu, 2018; Wu et al., 2021). The SFD around crater Zi Wei (C1) presented in Li and Wu (2018) are ideal for this purpose. Their measurements unequivocally demonstrate a real inflection point in the SFD shape, i.e., the SFD distribution does not continue monotonically along a power–law but has a convex up shape. The role of secondary fragments in forming the convex up shape can be excluded because of the following consideration. The modeled largest blocks produced by the Zi Wei crater of diameter 450 m is 21 m (Fig. 16 in Bart and Melosh, 2010). These blocks shall have produced background fragments starting at size <2 m (<10% size parent). The convex up shape of the SFD observed at this crater starts at 5 m with an inflection occurring between ~1 m and ~5 m (Li and Wu, 2018). Therefore, background daughter fragments do not contribute at this size range and cannot be responsible for the convex up shape. Although the current study consider only blocks above 1 m, it can be noted that a convex up shape is reported for sizes below 1 m for several locations: Surveyor VII (Shoemaker and Morris, 1970; Cintala and McBride, 1995), Station 4 of Apollo 16 (Muehlberger et al., 1972), Chand'E–3 (Di et al., 2016) and Chang'E–4 (Wu et al., 2018).

To calibrate the updated model, a measurement of block SFD at Tycho crater is performed. This population is ideal for the calibration because of its three properties. These properties make this population unique on the Moon.

1) The age of Tycho impact crater, and so the exposure age of its ejecta block, was determined using radiometrically dated samples associated with the impact (Drozd et al., 1977). It is to be noted that the link between the returned samples and the impact remains unconfirmed (Stöffler and Ryder, 2001). Nevertheless, the age of Tycho is at present one of the most accurate calibration point of the lunar chronology.

2) The age of 109 Myr (Drozd et al., 1977) is sufficiently old for the block SFD to have developed a mature shape while not having being completely erased.

3) The diameters of the blocks span a wide range of sizes (1–60 m) and are sufficiently large to be well detectable in LROC/NAC images.

The model calibration is performed as follow. Six model SFDs are calculated representing each combination of impact shattering function and meteoroid SFD. For each of the six model SFD, the detectability threshold ($Q/Q^*_s$) is determined through a fit to the measurement at Tycho that minimizes the RMS deviation. For this fit, the exposure age is known and parameters *a* and *b* are determined with a fitting procedure (see next section).

An independent test of this calibration is performed with the measurements of Basilevsky et al. (2013). That study reports the exposure time necessary for a reduction of the block population, of size 2–4 m, at a 50% and 99% level. The ages of ejecta block fields used in that study were based on the ages of the craters ejecting them. The ages of the craters themselves were estimated based on morphological criteria and, in few cases, based on the radiometric age of returned samples from the craters (Basilevsky et al., 2013).

**2.7 Fitting procedure and the initial block size–frequency distribution**

Once the input functions are known, a model SFD that matches a measured SFD is determined with three parameters: the age and two parameters defining the initial SFD at t=0 (*a*, *b*). The quality of the fit is measured with the Root Mean Square of normalized residuals (RMS). The calculation of the residuals starts at size above 1.4 m and each bin has equal importance. The age with the lowest RMS (best fit) is reported together with lower and upper age boundaries. We define these boundaries as the ages with 50% higher RMS than the best fit. There is ample evidence that the size–frequency of ejecta blocks formed by impact cratering follow a cumulative power law with slopes between -2 and -7 (e.g., Melosh, 1989; Bart and Melosh, 2010; Kumar et al., 2014) consistent with the slope measured for the cumulative SFD of secondary impacts (e.g., Bart and Melosh, 2010; Shoemaker et al., 1965). A power law distribution is adopted as initial SFD and is defined by two unknowns: the proportionality constant (*a*) and the power law exponent (*b*). The frequency of the few largest blocks does not often follow the power law (e.g, Bart and Melosh, 2010; Li, B. et al., 2017; Krishna and Kumar, 2016) and are not considered in the fitting procedure. We note that in reality some ejected blocks, in particular the small ones, might have had a previous history of exposure (e.g., Burnett et al., 1975).

**2.7 Measurement requirements**

It is important to recognize that the properties of the power law (parameters *a* and *b*) describing the initial block SFD shape change spatially around a source crater. This spatial variation is exemplified in the study of Krishna and Kumar (2016). That study revealed the presence of major spatial trends: the abundance decreases away from the rim, the exponent slightly decreases (steeper slope) away from the rim, and the abundance is higher within crater rays. For meaningful comparison between model and measured SFD, block counts need to be performed only outside the crater, and within areas that display a random distribution of blocks. For example, only within a ray, or only near the crater rim, and not in an area combining near and distal block fields. Formation of block fields by other process (e.g., mass wasting at ridges or within the crater) might lead to a different initial SFD and must be avoided.

The size of the count area should be large enough to contain a wide range of block diameters that are needed to reconstruct the initial SFD. We can expect that for very small areas, the stochastic process of impact bombardment might play a role such that if a block field is formed by a single event in time, the ages estimated from different areas within that block field can vary and be slightly higher and lower than the actual formation time. The quantification of this effect is left for future studies.

The spatial density of blocks in the count area is irrelevant because the spatial density, i.e., the initial SFD distribution, is a free parameter determined together with the age. In order to cover a wide range of block diameters, it is nevertheless preferable to choose a high density area. For young block fields, however, it is practically useful to choose a low spatial density area where a complete count of very small blocks (~1 m) can be achieved. High density areas will contain a very high number of ~1 m blocks that is challenging to count (e.g., the underlying regolith in high density area is not visible at Giordano Bruno crater). Finally, we note that counts are performed on images with the smallest possible ground sampling dimension and with moderate incidence angles (~40º-70º) to facilitate relief identification.

**3. Results**
**3.1 Overall model SFD shapes**

The shattering energy functions and the projectile SFDs strongly influence the shape of the block SFD and its changes with time (Figure 4). As expected, the shattering function of BA99 (Figure 4a,b,c) requires more energy for shattering than the function of HH99 (Figure 4d,e,f) and so it relatively hampers the decrease of block abundance with time. The rather small differences in the projectile SFDs (Figure 3) have great influence in the block SFD shape. The steeper projectile SFD leads to enhanced decrease of block abundance, in particular at small sizes. In general, the SFD shape is influenced by the initial SFD (t=0). For the BA99 with D17 model, a lower initial power–law index has the effect of favoring a convex up shape, i.e., the bending is more pronounced (shallower) and occurs at younger ages. The size of the smallest survivor is independent of the initial power law exponent, however. Intuitively, and as per how the model is constructed, the convex up shape does not depend on the proportionality factor of the power–law. It is found that despite the high flux of lunar secondaries, their impacts minimally increase the erosion rate because their velocity is insufficient in shattering the coherent blocks. For impact velocities in the range 0.5–1.5 km/s, the destruction rate for 1 m sized block is 0.04% after 500 Myr and is lower for larger blocks.

The stochastic nature of meteoroid bombardment considered in the model does not change significantly the SFD shapes. In fact, several model runs with the same input functions produces SFD shape that are undistinguishable between each other.

The SFD shapes reported in Figure 4 are calculated with an arbitrary detectability threshold that divide a population into intact and "erased" blocks. This is a crude subdivision because blocks accumulate the energy imparted by impacts in a gradual fashion (Figure 5). The energy threshold above which blocks are sufficiently shattered to be undetected in LROC/NAC images is determined in the following section.

**3.2 Model calibration**

In order to determine appropriate values of the detectability ("erasure") threshold, the model SFD are compared to measured block SFD at Tycho crater. The BA99 function requires rather small values of $Q/Q^*_s$ (<2), whereas the HH999 function necessitate higher values $Q/Q^*_s>7$. The meteoroid SFD B02 has the highest RMS deviation, is not able to match the largest data bins and requires the smallest possible power law exponent. This poor fit is likely due to the characteristic power law shape (Figure 6a,d) resulting from this meteoroid distribution SFD. Fits with lower RMS are obtained with the meteoroid SFD G85–B02 and D17. The meteoroid SFD G85–B02 underestimates the abundance of small blocks with both shattering function (Figure 6b,e). This is not expected because the block measurement at small sizes is likely to be underestimated rather than overestimated. The models D17 with BA99 and with HH99 show the smallest RMS of all models and reproduce the SFD across the entire range of diameters. The meteoroid distribution D17 (and G85–B02 to a lesser extent), independently of the shattering function, can reproduce both a power–law, for young ages, and a convex up shape, at later times (Figure 4c,f). This result strongly suggests that these projectile SFDs well characterize the average meteoroid SFD over many million years.

Among the six possible models, the exposure age estimates with the meteoroid SFD of D17 fall the closest to the measurements of Basilevsky et al. (2013) (Figure 7). The updated models with D17, however, overestimates the age at which 50% of a population should be erased, whereas it can reproduce the exposure age for a 99% reduction of the population. We remark that Figure 5 is the updated version of Figure 10 in Hörz et al. (1975).

For the remaining of this article, the following input functions are used because they best approximate the measurements at Tycho: Shattering function of BA99, meteoroid SFD of D17, detectability threshold at 1.69Q*s.

We present a plot comparing the survival time versus target size (Figure 8). The survival time is arbitrarily defined at 90% of the population being above the detectability threshold.

**3.3 Age of lunar ejecta block fields**

We performed block SFD measurement in a low spatial density area near the rim of Byrgius A crater (Figure 9). For this young block field, blocks as small as ~1 m were carefully counted and used in the fitting procedure. The best fit age is $27^{+40}_{-27}$ Myr (Table 1).

In order to assess the precision in age calculation due to the unknown parameters *a* and *b* of the initial power law, the fitting procedure was applied again to the Tycho block counts, this time with the detectability threshold fixed and the age as third free parameter. The best fit age is $120^{+71}_{-51}$ Myr (Figure 10). The best fit age is not 109 Myr, as set by the calibration, because of the use of a slightly different sampling of the parameter space of variables *a* and *b*.

We performed block SFD measurement near the rim of Aristarchus crater. We find that the exposure age providing the best fit to the measurement is $207^{+61}_{-51}$ Myr (Figure 11). The SFD presents a convex up shape that is due to the mature population and is not the result of an image resolution bias.

We performed block SFD measurement near the rim of Jackson crater (Figure 12). The best fit age is  $298^{+1}_{-1}$ Myr. Again, the SFD display a convex up shape that is not due to an image resolution bias.

We performed block SFD measurement near the rim of Copernicus crater and determined a best fit age of $650^{+50}_{-100}$ Myr (Figure 13). The initial power law index (*b*) is determined with less reliably than at the other locations because the data points at the largest diameters are too few and the best fit exponent is the maximum we allowed (-2). This is in contrast to the measurement at Aristarchus and Jackson and is due to the very mature block field of Copernicus (Figure 14). We find that the ages cover the range 650-900 Myr for exponent values in the range -2<b<-4. The data point at diameter 63 m is not reproduced because it is the largest diameter bin of the model and it is artificially excluded.

**3.4 Age determinations at Jackson crater using crater SFD**

For comparison purposes, we performed crater SFD measurements at several locations around Jackson crater and derived absolute model ages with a crater chronology function (Neukum et al., 2001) and a crater production function (Neukum et al., 2001). Four areas were studied: three on the ejecta and one on a melt pool (Figure 15). In a large area of 1x10$^3$ km$^2$ on the ejecta the age calculated with crater diameters in the range 200-700 m is $140^{+9}_{-9}$ Myr. In a small area of 0.6 km$^2$ on the ejecta the age calculated with crater diameters in the range 10-80 m is $154^{+6}_{-6}$ Myr. In another small area of 0.5 km$^2$ on the ejecta the age calculated with crater diameters in the range 20-70 m is $210^{+20}_{-20}$ Myr. In an area of 2.7 km$^2$ on a melt pool the age calculated with crater diameters in the range 10-80 m is $85^{+4}_{-4}$ Myr.

**4. Discussion**
**4.1 Survival times of blocks**

The plot presenting the survival times (Figure 8) is the same as Figure 11 of Hörz et al. (1975), including now larger diameters. The survival times for targets of about 3 cm in size are the same for both old and updated model. For diameters larger than ~5 cm the destruction rates are ~2 times higher than in the original model, i.e., survival time are shorter. This difference is however within the range of uncertainty on the crater production rates recognized in the original model. The crater production rate was used in the original model instead of the meteoroid mass flux and was estimated to be known within a factor of 5. Specifically, the faster destruction times of the updated model are consistent with the highest, hypothetical limit on crater frequency, referred to as "rate III" in Hörz et al., 1975. Alternatively, the difference could be explained by large rocks being more strongly shattered in the updated modeled. This could explain why the two models agree on the destruction rate for a rock size of ~3 cm. A higher destruction rate at larger sizes could be explained by the shape of the meteoroid SFD, which was not known at the time of the original model, and/or by the shattering functions, that relative to the shattering function used in the original model, favor shattering of large rocks.

The slight increase in survival time below ~5 cm was not expected (Figure 8). To calculate such values the SFD of meteorites as small as $1\times10^{-4}$ m ($1\times10^{-9}$ kg) was used. The SFD at these small diameters is relatively well constrained (Grün et al., 1985). The specific energy function is also reliable down to 1–2 cm target size based on impact experiments performed at such small sizes (e.g., Housen and Holsapple, 1999). Therefore, the increase of survival time for rocks smaller than 5 cm is considered reliable and likely due to the decrease in steepness of the meteoroid SFD for projectile less than about $1\times10^{-3}$ m ($1\times10^{-6}$ kg), i.e., relatively fewer very small impactors (Grün et al., 1985). This slight increase in survival time is consistent with the observations at South Ray crater (2 Myr in age) of increased abundance of rocks of size below about 10 cm (Muehlberger et al., 1972). Without the increase in survival time, almost all (90%) of rocks a few cm in size should have been erased after 2 Myr of exposure time.

The higher destruction rates of the updated model with respect to the original study confirm, although do not exactly match, the findings of Basilevsky et al. (2013) and Ghent et al. (2014). The age calculated to destroy 50% of an initial population is larger than the estimate of Basilevsky et al., (2013), whereas the time necessary to reach 99% destruction agrees (Figure 7). The mismatch could be due to the estimate of exactly 50% decrease in rock density based on images. Such estimate needs to assume a uniform initial spatial rock density in all investigated areas and images. This might not have been the case. In addition, the age of some of the block fields used in Basilevsky et al. (2013) might be inaccurate due to the difficulty in estimating a crater age (and its ejecta block age) from morphology or from the crater age dating technique in areas of low crater density.

The studies of Ghent et al., (2014) and Wei et al. (2020) relate the rock abundance from Diviner thermal measurements and Chang'E–2 microwave to the exposure age of craters. A detailed comparison with the updated model is not straightforward because the initial rock abundance at such craters is not known. It can only be said that the updated model and their measurements agree for ages in the 80-500 Myr range. A comparison at younger ages cannot be done because the data was not available in the study of Wei et al. (2020) and was based on a single measurement in Ghent et al. (2014). Deviation at older ages is expected between the updated model and the two studies because the measured rock abundance reported in the two studies is the result of recent small impacts and newly excavated blocks rather than the pristine population. In addition, the relationship of Wei et al. (2020) is based on exposed as well as buried rocks, whose evolution is different from the updated model.

**5.2 Age estimations using block SFD**

The block count at Byrgius A crater clearly points to an age below 50 Myr and is consistent with the crater SFD measurements reported by Morota et al., (2009) of $47^{+14}_{-14}$ Myr. The comparison between the ages from blocks SFD with the ages from crater SFD at Aristarchus and Jackson crater is not straightforward because crater SFD ages on ejecta blankets record the surface exposure time as well as the amount of self-secondary craters (Zanetti et al., 2017). There is indeed a wide range of possible ejecta ages for Arisarchus and Jackson. At Aristarchus the ages reported in the literature using the crater dating technique are 130–180 Myr (König and Neukum, 1976), 164±1.4 Myr or ~280 Myr (Zanetti et al., 2017). Our block SFD age falls within this wide range. For Jackson crater, the range of our crater SFD ages is from $140^{+9}_{-9}$ to $210^{+20}_{-20}$ Myr whereas the best fit age with block SFD is older. If we assume the crater SFD ages to be correct, a hint that the block SFD age is inaccurate might be given by the RMS distribution (Figure 14). Figure 14 shows the minimum RMS for Jackson is not found within the broader low RMS area. This feature is not found at the other measured locations. The investigation of this behavior is left for future studies.

The range of ages estimated for the block field of Copernicus crater (650-900 Myr) cover the age of the Copernicus impact event determined from Apollo samples and estimated to be 800±15 Myr by Stöffler and Ryder (2001). There is a second, less reliable, approach that can be taken to estimate the age of Copernicus block field. It is clear from the plateau shape of the measured SFD (Figure 14) that the smallest survivor blocks in the count area are in the size range 10–20 m. The 90% destruction time of 10–20 m blocks is 500–800 Myr (Figure 8), partly overlapping the range determined by fitting the entire SFD. This second estimate is not reliable, however, because it is based on an arbitrary definition of erasure (here 90%) and the initial abundance is not considered. In addition, the used range of smallest survivors is only valid within the count area and it is not excluded that other areas around Copernicus rim host smallest survivor blocks of slightly larger/smaller size that have/have not been destroyed by the stochastic process. Thus, the survival time values (Figure 8) are useful as theoretical information and should be treated with caution when comparisons are made with observations because usually, unless a SFD is measured, the initial abundance is not known. Survival time is a theoretical value that does not necessarily correspond to the exposure age of a rock field. It provides, however, a first order estimate of the age of a block field.

### 5.3 The role of thermal fatigue and similarity with block SFD on Mars

Can the role of thermal fatigue in lunar block erosion be excluded based on the results presented above? The variety of possible input functions are likely preventing to draw conclusions on this aspect. Let's assume that the meteoroid SFD follows the function of B02. In order to remove preferentially small (few m) blocks and match the observed SFD and the characteristic convex up shape, the shattering energy should decrease with decreasing sizes, i.e., opposite trend to the used shattering energy functions. This effect should be explained by a decrease of block strength due to thermal fatigue. To reproduce the observed SFD, this decrease in shattering energy would need to occur at increasingly larger sizes with increasing exposure age. Thus, the convex up shape explained with the D17 and B02-G85 functions could be mimicked with the B02 function and an hypothetical block weakening function to represent the effect of thermal fatigue. However, it is not clear whether internal fractures known to develop by diurnal temperature variations (Delbo et al., 2014; Molaro et al., 2017) would lead to an overall weaker block, or the contrary, to a more resistant block upon impact bombardment due to an increased porosity (e.g., Jutzi et al., 2010; Flynn et al., 2015). In addition, it seems to be unlikely that block strength decrease with time due to thermal fatigue has an effect on block SFD very similar to that played by a meteoroid SFD.

As final note, we point out that the model SFD with the characteristic convex up shape that develops at moderate to old ages best reproduce the SFD measurements on the Moon and, surprisingly, provides a good match to SFD measurements on the surface of Mars (e.g., Fig. 17 in Golombek et al., 2008). This latter similarity might be a coincidence or suggest that different geological processes produce the same type of SFD shape (e.g., Craddock and Golombek, 2016). Alternatively, and less likely, the similarity might hint that the surface of Mars was gardened by cm to m-sized meteoroids at some point in the past when the atmosphere and its protective effect were strongly reduced. The prove or discount of this far reaching implication is behind the scope of this study.

### 5. Conclusions

The model of block catastrophic rupture of Hörz et al. (1975) is revisited in light of new understanding of the functions describing the energy necessary for block shattering and improved estimates of the flux and size–frequency distributions (SFDs) of meteoroids hitting lunar blocks. The input functions that best reproduce the number and size–frequency distribution of blocks on the lunar surface are identified. With such functions, the modeled block SFD well reproduce measurements of block size–frequency distributions in the range 1 m to few 10s of m. The outcomes of this study are:

- The evolution of lunar block SFD in the range 1–50 m is as follow: For young (<~50 Myr) population, the SFD is best approximated by a power law, whereas for older populations, the extrapolation at small diameters is best performed by an exponential distribution. The age at which the exponential distribution starts to develop decreases with decreasing steepness of the initial SFD. In all cases, the size of the smallest observable fragment is proportional to the exposure time.
- The calculated survival times of rocks as a function of size from 1 cm to 50 m represent the most accurate estimate to date. For rocks above ~5 cm, the survival time increases with increasing size, whereas for rocks below ~5 cm, the survival time slightly increases with decreasing size.
- The flux and SFD of meteoroids in the size range 0.01–1 m measured nowadays is a good match to the average SFD of meteoroids recorded over hundreds of million years by the SFD of mature block populations.
- Surface exposure age of block fields can be estimated from SFD measurements.

## 6. Acknowledgment

OR, RMM, and MP are supported by a Sofja Kovalevskaja award of the Alexander von Humboldt foundation. The authors are grateful to Friedrich Hörz for a discussion of this study. This article is dedicated to Alexandra Elbakyan.

## 8. Data and code availability

LROC/NAC image data are available at http://wms.lroc.asu.edu/lroc/search.

The code for the updated model is available upon request to Ottaviano Rüsch.

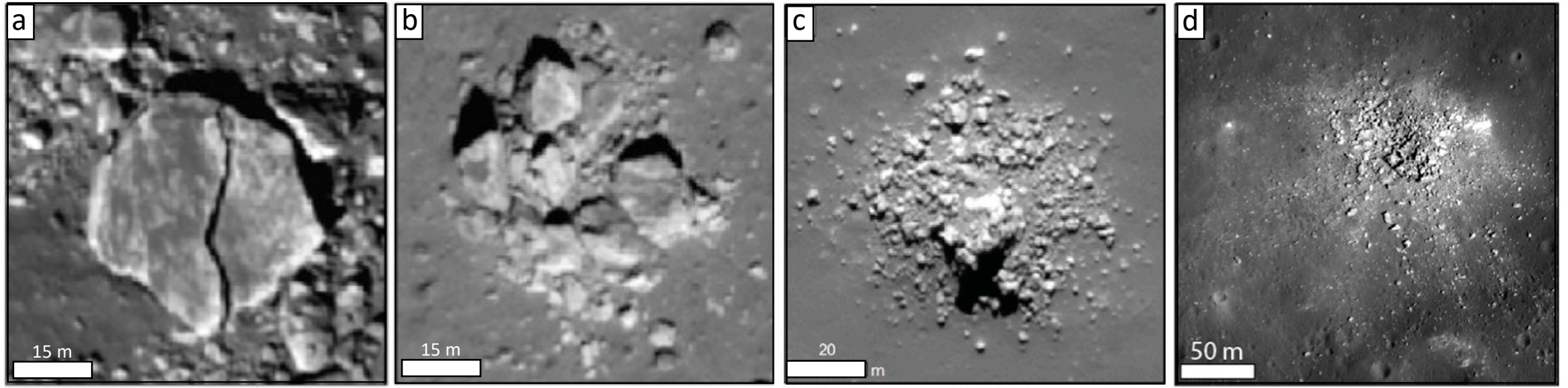

**Figure 1.** Examples of lunar blocks likely shattered by impacts observed with the LROC NAC camera. (a). Fractured block. (b) Shattered block with few large fragments. (c) Shattered block with remnant core. (d) Highly shattered block with the size of the largest fragment much smaller than original block. The distant, smallest fragments in (d) are undistinguishable from the primary block population. All images from Ruesch et al., (2020).

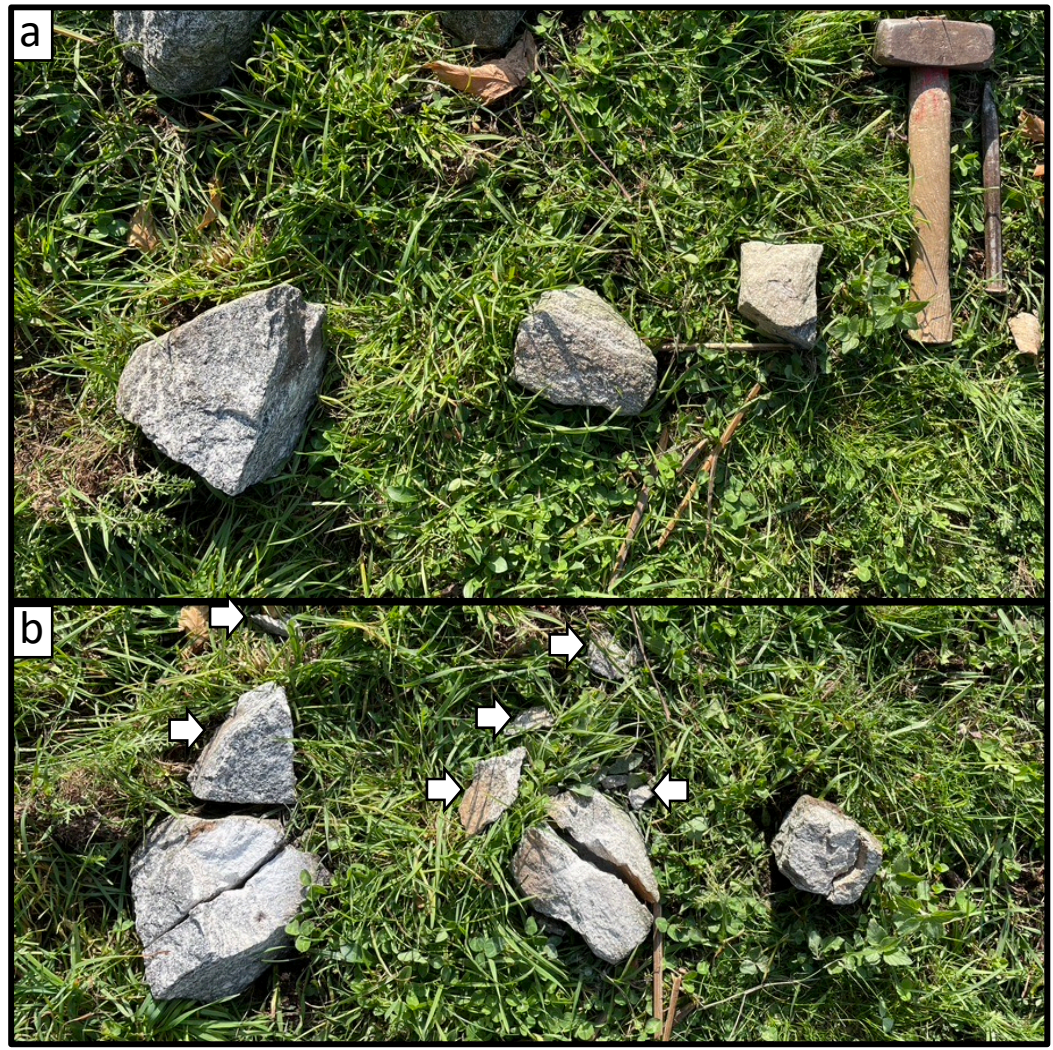

**Figure 2.** Simple experiment demonstrating that the largest fragments of shattered rocks resting on soil are not moving far from their original location, and this is independent of size. Smaller spall fragments, instead, move slightly away (arrows). The small, medium and large gneiss rocks were hammered with a point 6, 60 and 210 times, respectively, until $M_{lt}/M_t \sim 0.5$. Panels (a) and (b) show the initial and final state, respectively. Hammer length is 25 cm.

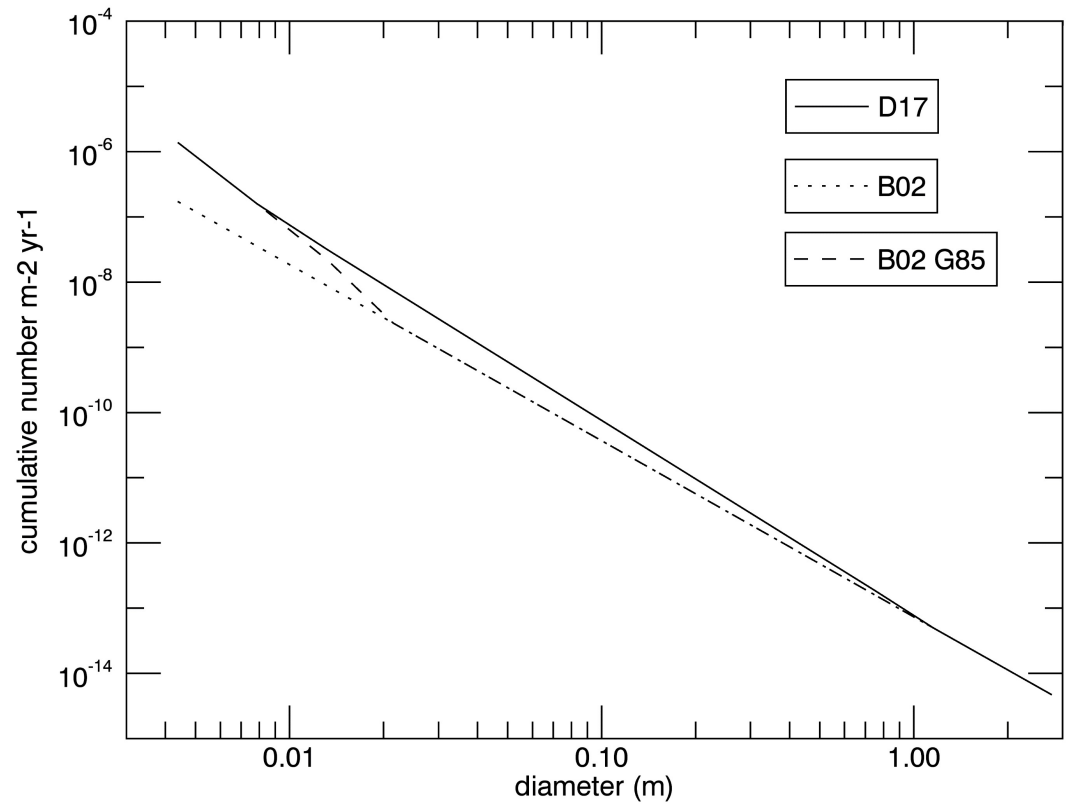

**Figure 3.** Three estimates of the size-frequency distributions of projectiles hitting the surface of the Moon from the studies of Grün et al. (1985) (label G85-B02), Brown et al. (2002) (label B05) and the interpolation of Drolshagen et al. (2017) (label D17). All three distributions are presented and discussed in Drolshagen et al. (2017).

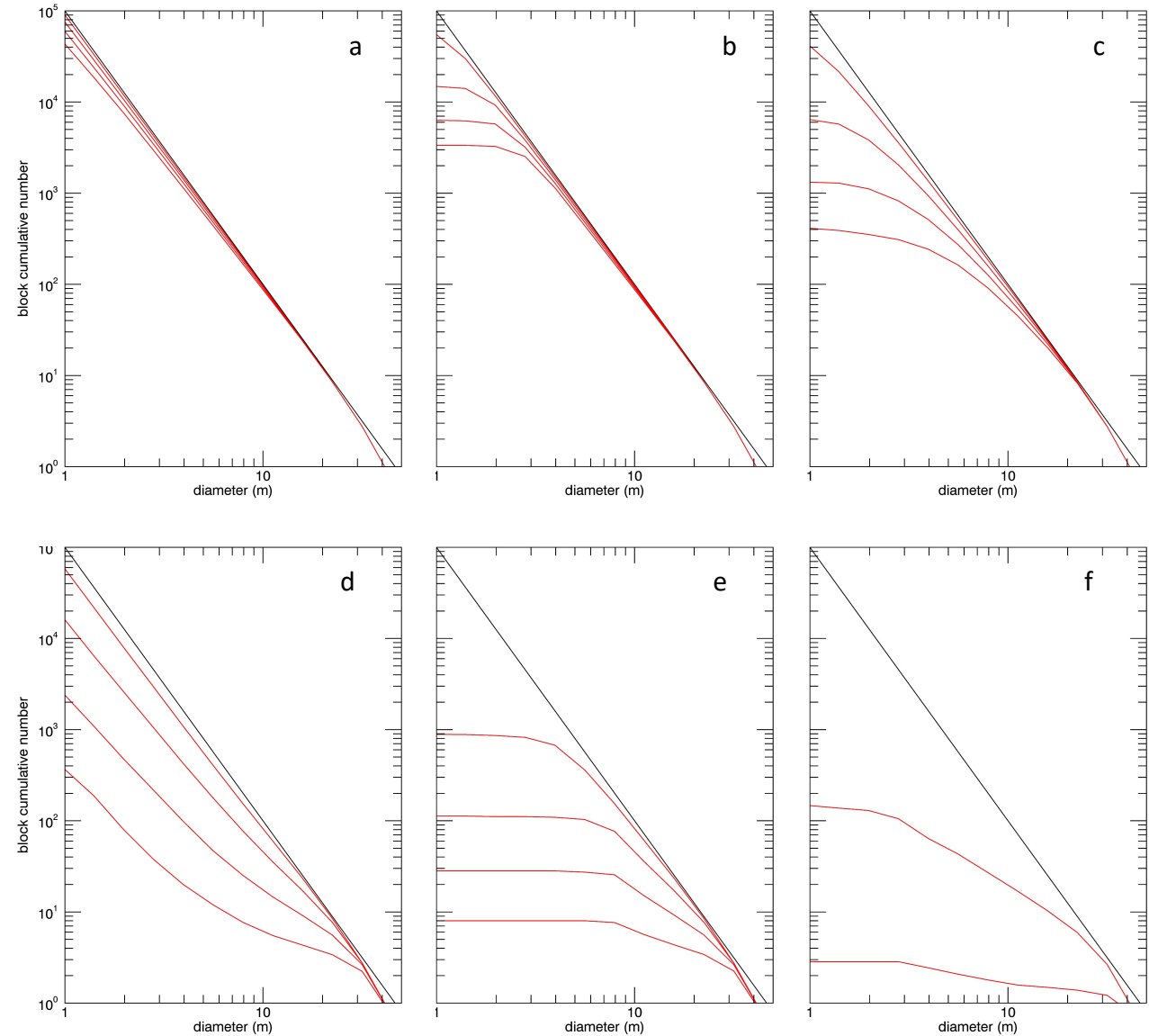

**Figure 4.** Evolution of the size-frequency distribution (SFD) of blocks for different projectile SFDs and different functions describing the energy for block shattering. The initial block SFD is shown in black with a power law index of arbitrary value -3 for illustration purposes. Red lines are isochrones at 50, 100, 150, 200 Myr of surface exposure time. Panels (a,b,c) calculated with the Benz and Asphaug (1999) shattering function. Panels (d,e,f) calcualted with the Housen and Holsapple (1999) shattering function. Panels (a,e) use the projectile SFD from Brown et al., (2002), panels (b,e,) the Brown et al. (2002) including the Grun et al. (1985), and panels (c,f) the Drolshagen et al. (2017) distribution. For illustration purposes, here the non detectable („erased") blocks are defined by a total accumulated energy $Q/Q^*_s=1$.

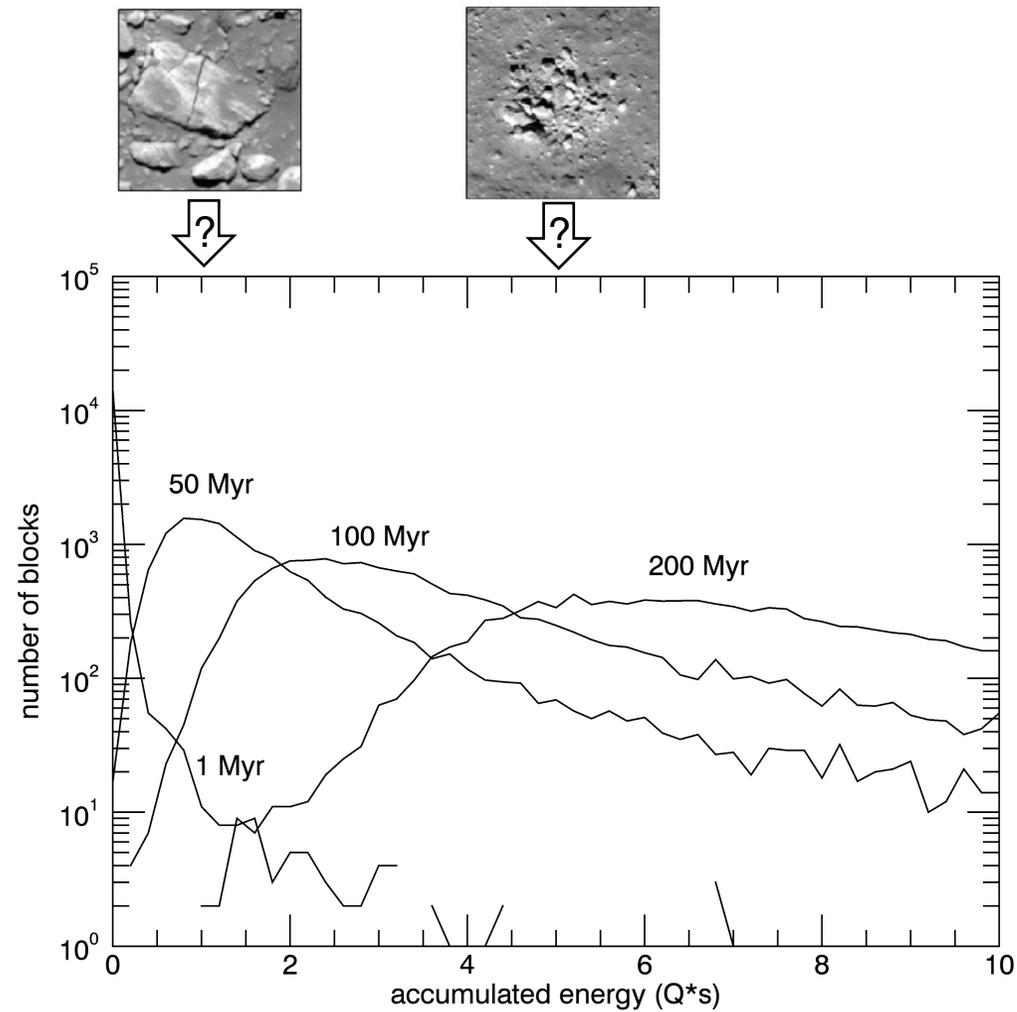

**Figure 5.** Histogram of the total energy accumulted by blocks as a function of surface exposure age. The energy is expressed with the specific energy $Q^*_s$. The model outputs are calcualted for 1 m sized blocks using the BA99 shattering energy function and the D17 meteoroid size-frequency distribution. Images illustrate the possible morphology of shattered blocks at different imparted energies (extracted from Ruesch et al., 2020).

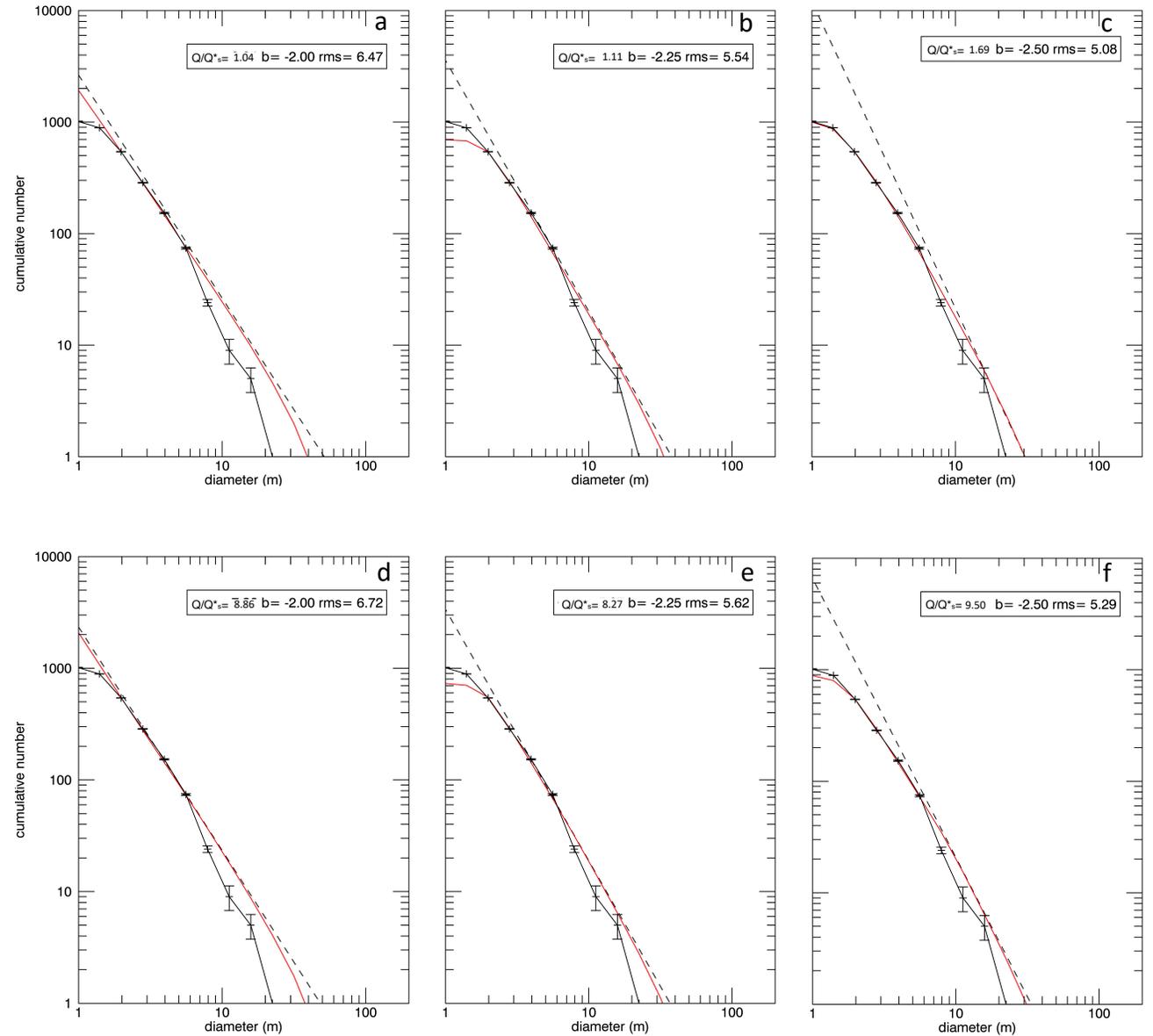

**Figure 6.** Comparison between measured block SFD at Tycho crater and modeled SFD using different input functions as in Figure 4. The detectability („erasure") threshold (Q/Q*s) and the slope of the initial SFD (b) providing the best fit to the data are shown. RMS calculated from 2 m to 20 m. The initial block SFD is shown as a dashed line. The model surface exposure time is set constant at 109 Myr. Panels (a,b,c) calculated with the Benz and Asphaug (1999) shattering function. Panels (d,e,f) calcualted with the Housen and Holsapple (1999) shattering function. Panels (a,e) use the projectile SFD from Brown et al., (2002), panels (b,e,) the Brown et al. (2002) including the Grun et al. (1985), and panels (c,f) the Drolshagen et al. (2017) distribution.

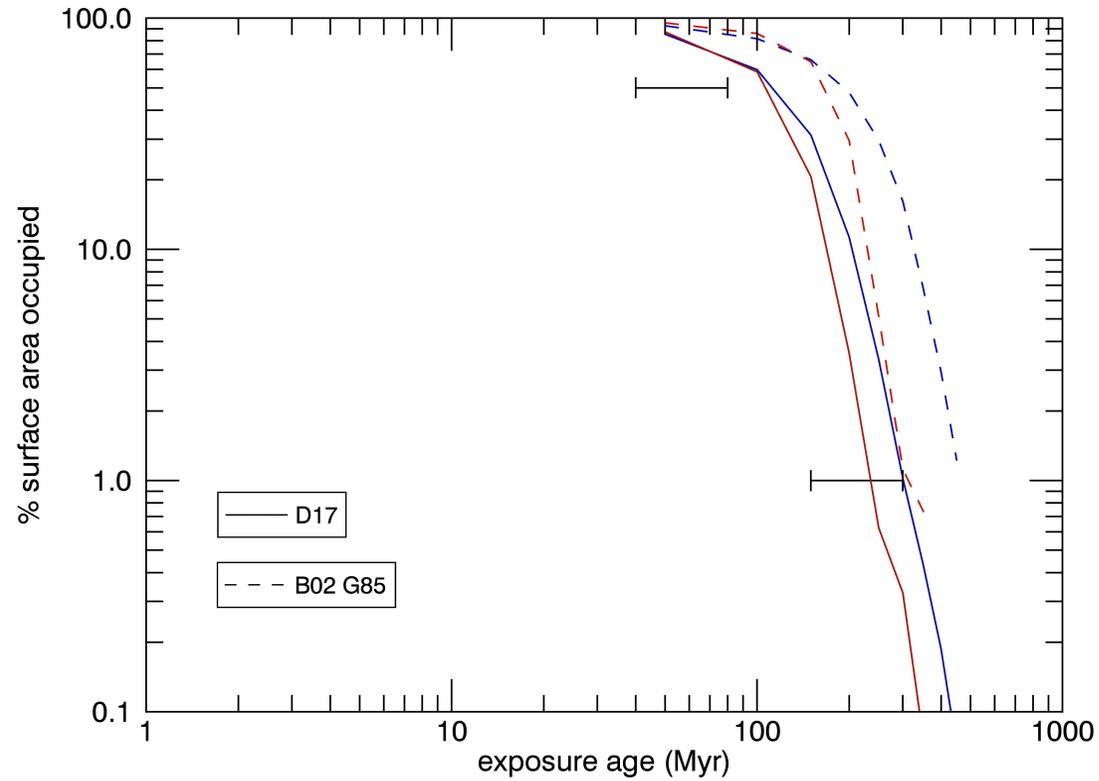

**Figure 7.** Decrease of the surface block coverage with time for blocks 3 m in size and for different input functions. The line styles denote different projectile SFDs as in Figure 1. Red and blue curves are calculated with the Benz and Asphaug (1999) and Housen and Holsapple (1999) functions, respectively. Horizontal bars are estimates based on observations from Basilevsky et al. (2013) for blocks 2-4 m in size. The detectability („erasure") threshold is calibrated and the same as in figure 6.

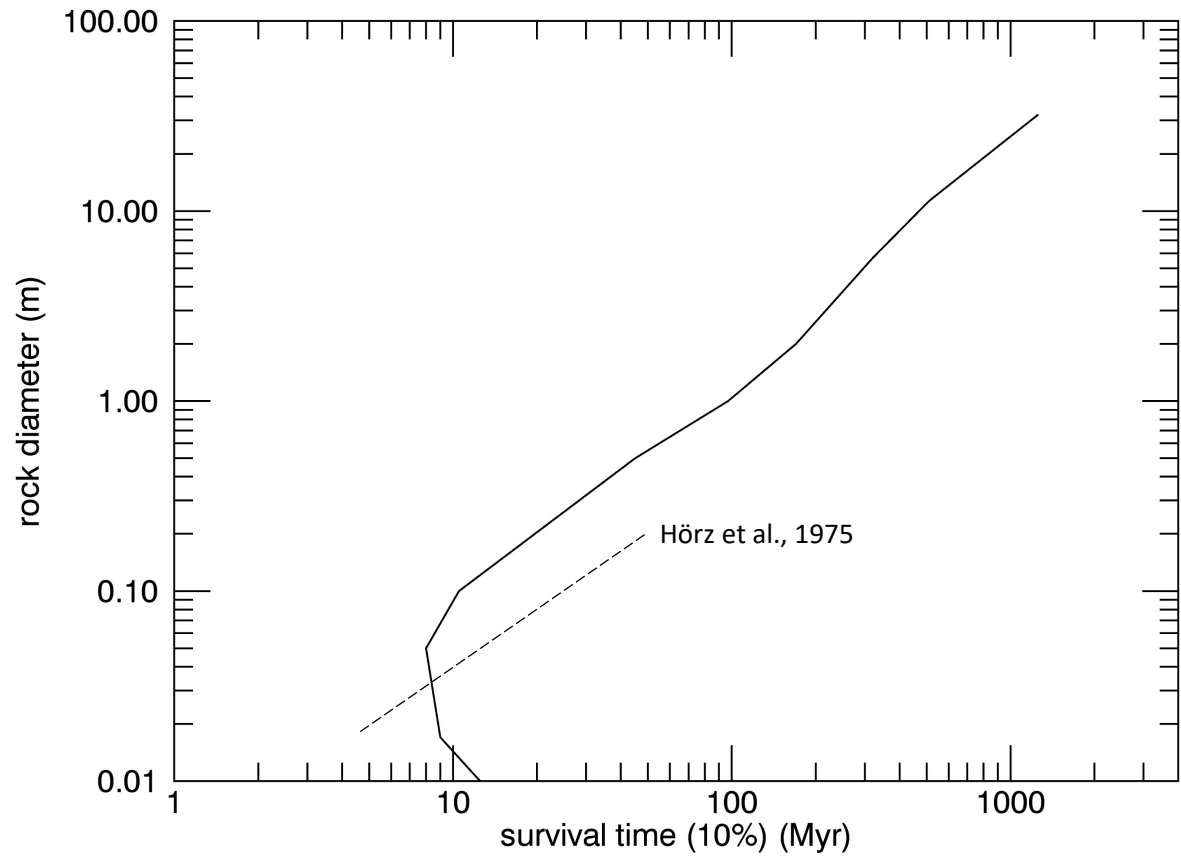

**Figure 8**. Survival time defined as the time necessary to catastrophycally shatter a rock of given diameter with a 90% probability. The dashed line is the estimate of Hörz et al., 1975.

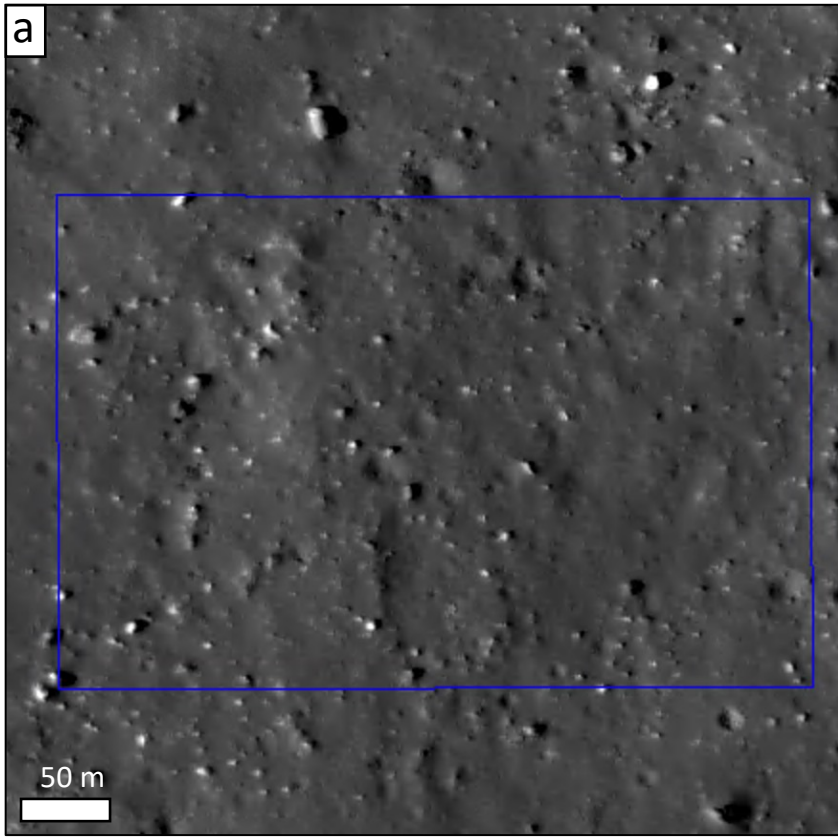 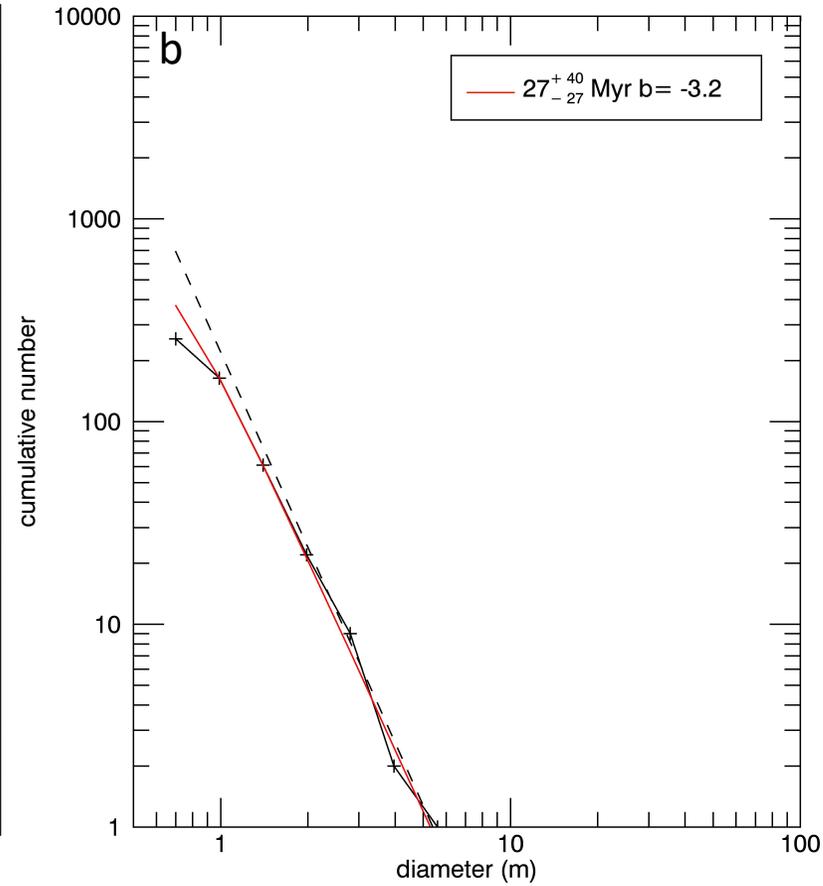 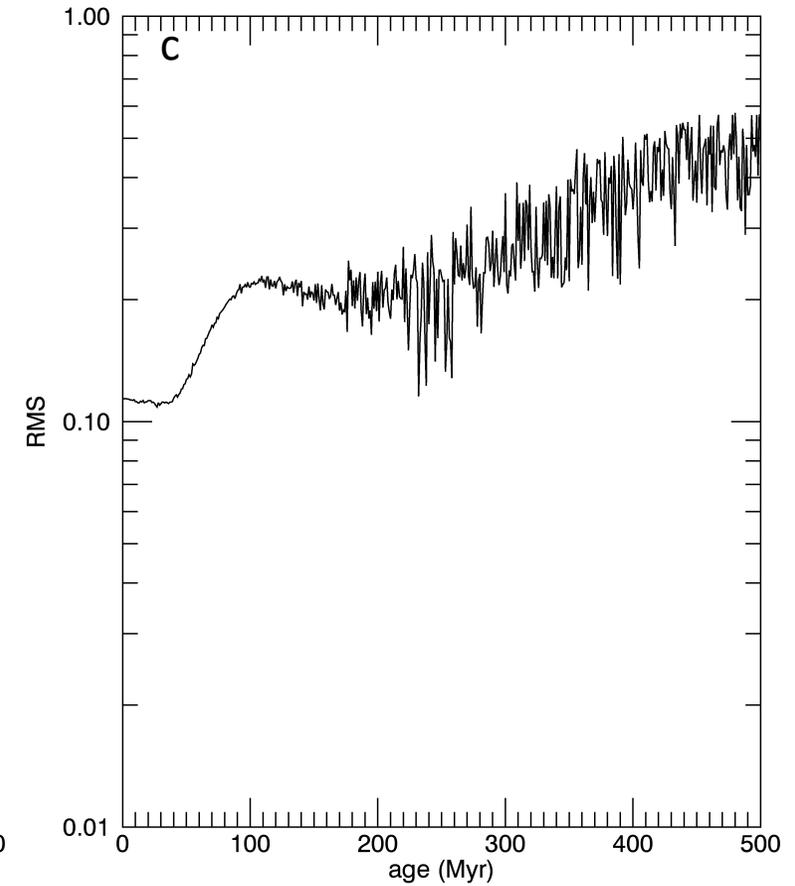

**Figure 9**. (a) NAC image M1265236990RC showing the block count area near the rim of Byrgius A crater. (b) Cumulative size-frequency distribution of blocks. Data shown with crosses, model initial SFD as dashed line and best fit isochrone in red. (c) RMS deviation as a function of exposure age.

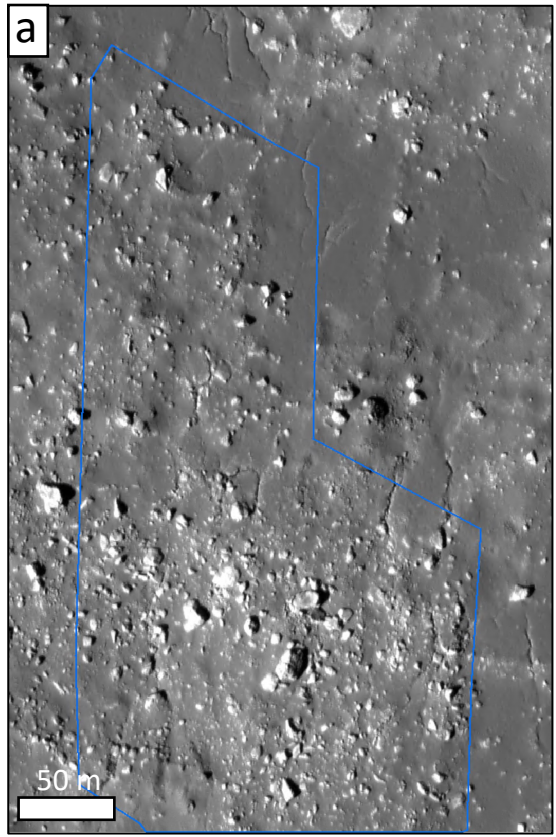 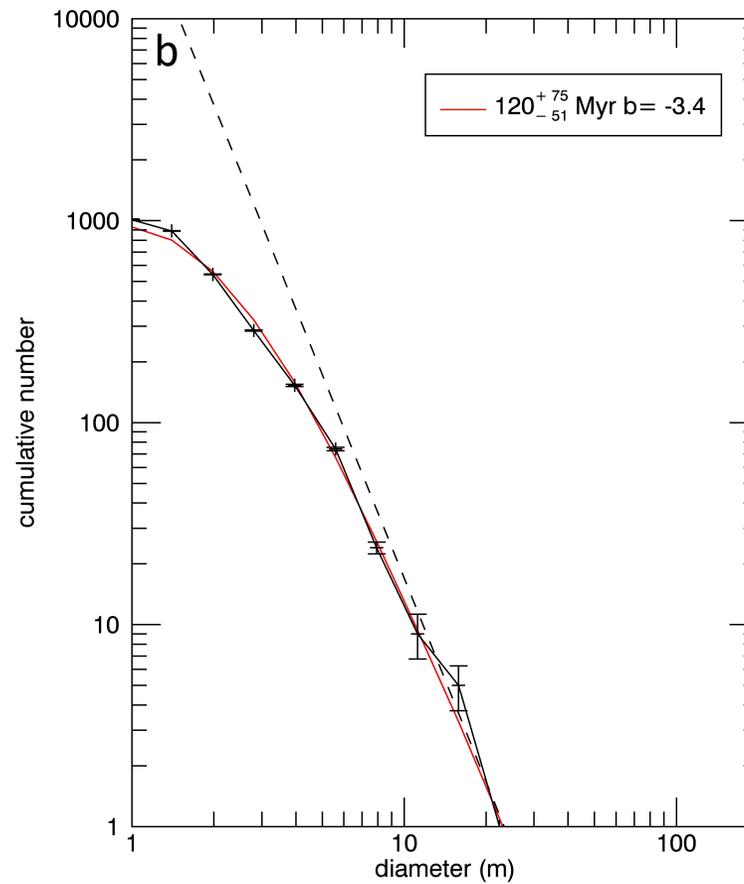 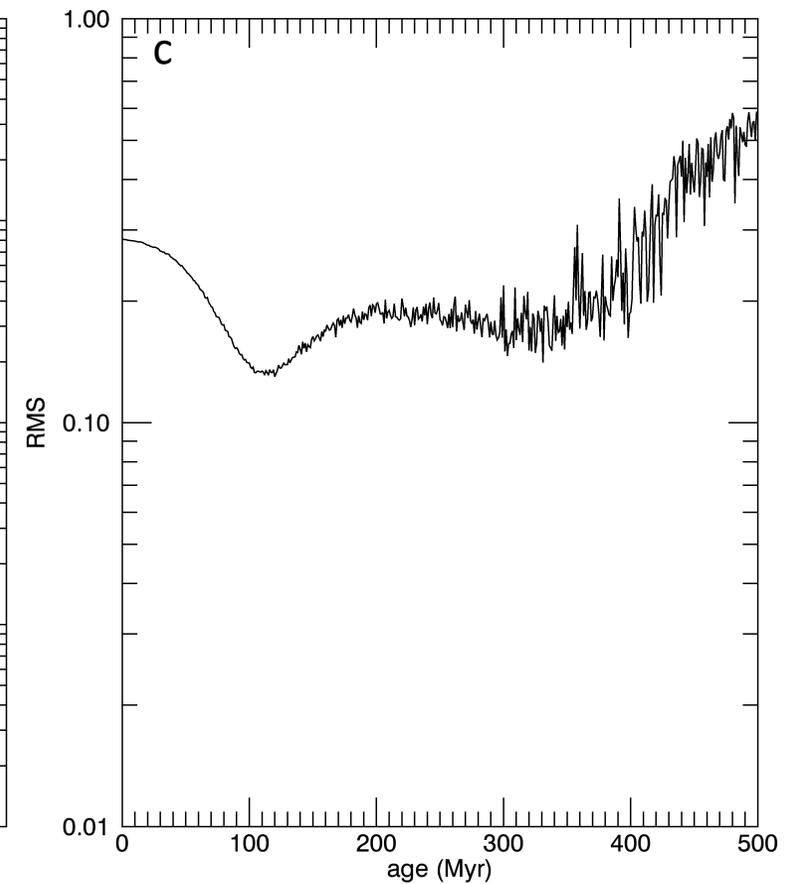

**Figure 10**. NAC image M1162548409RC showing the block count area near the rim of Tycho crater. (b) Cumulative size-frequency distribution of blocks. Data shown with crosses, model initial SFD as dashed line and best fit isochrone in red. (c) RMS deviation as a function of exposure age.

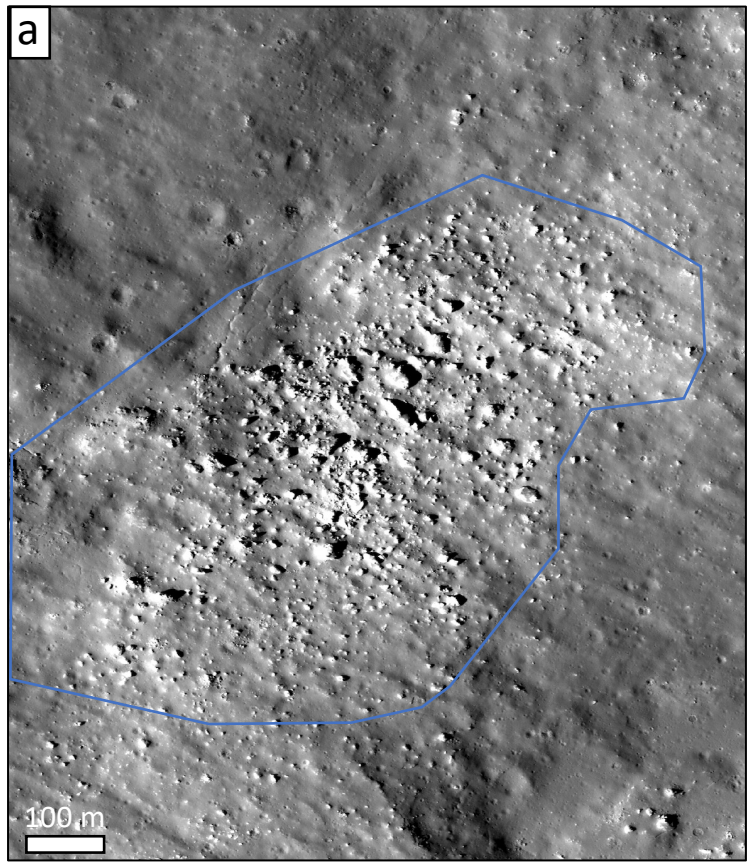 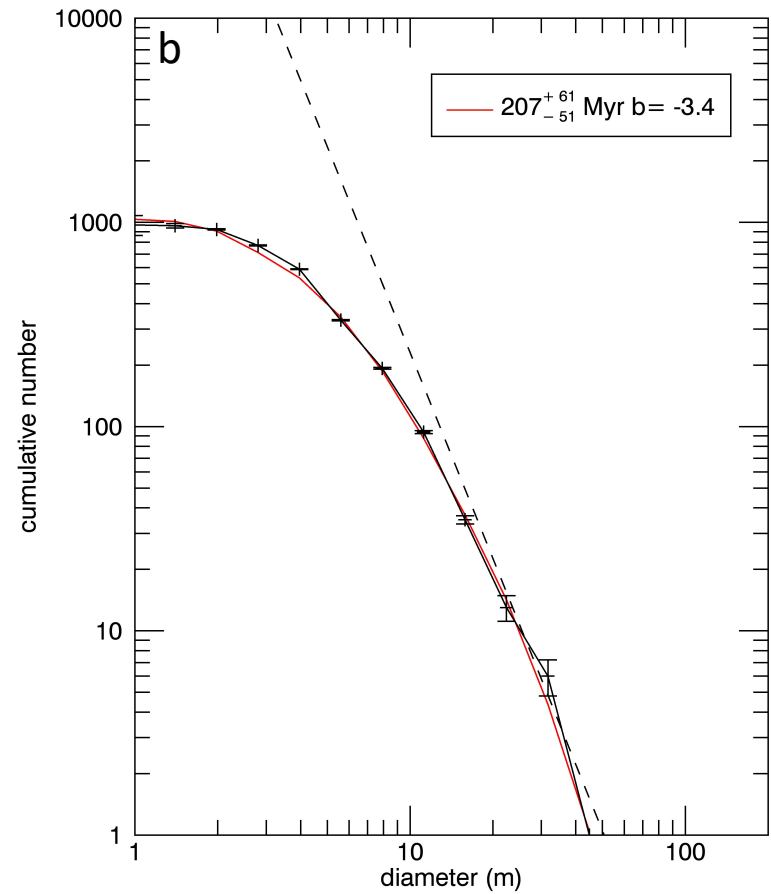 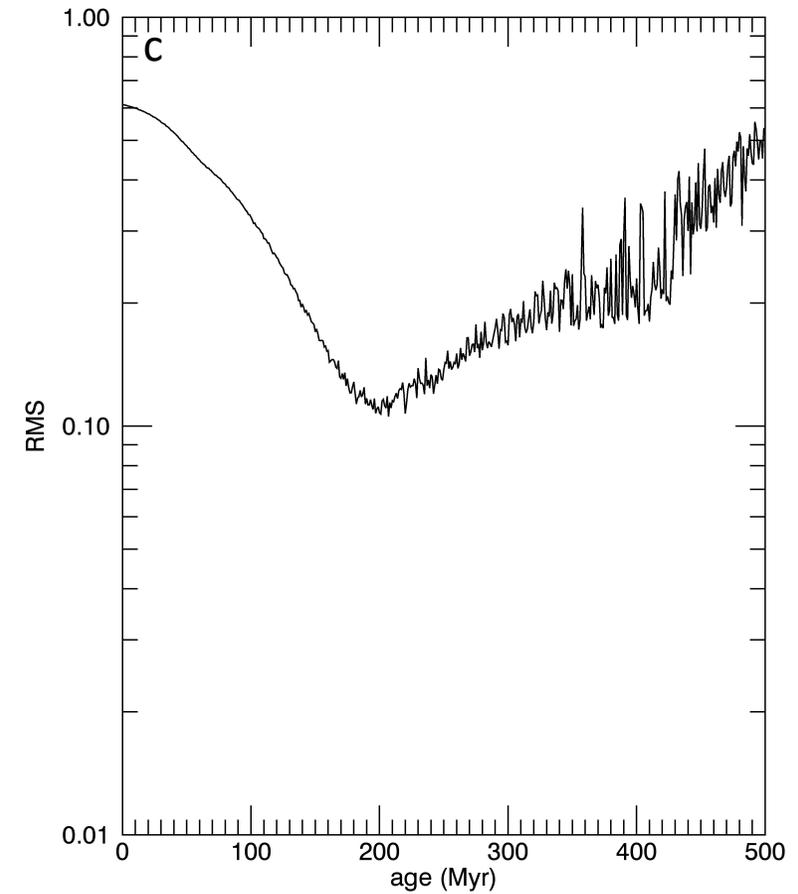

**Figure 11**. NAC image M144931504LC showing the block count area near the rim of Aristarchus crater. (b) Cumulative size-frequency distribution of blocks. Data shown with crosses, model initial SFD as dashed line and best fit isochrone in red. (c) RMS deviation as a function of exposure age.

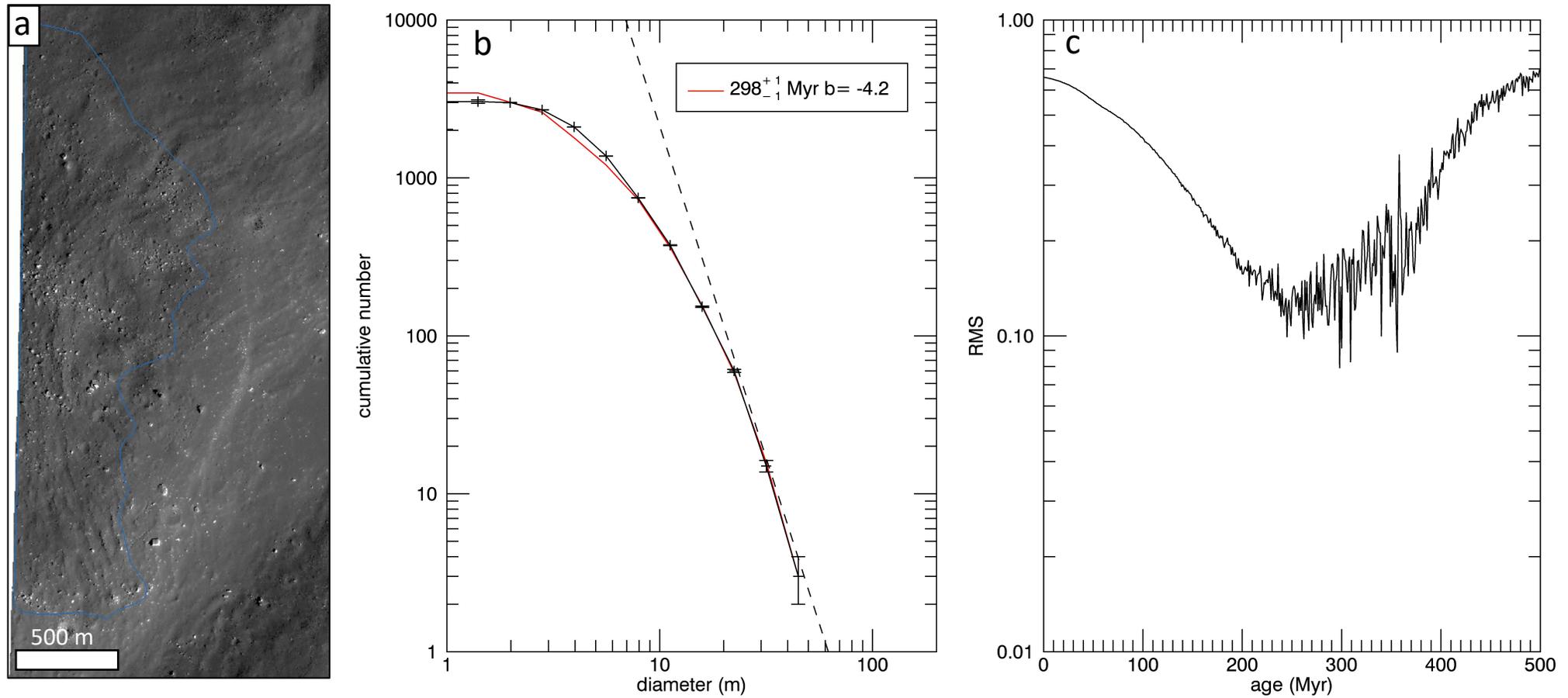

**Figure 12**. (a) NAC image M112671946RE showing a large area of block count near the rim of Jackson crater. (b) Cumulative size-frequency distribution of blocks. Data shown with crosses, model initial SFD as dashed line and best fit isochrone in red. (c) RMS deviation as a function of exposure age.

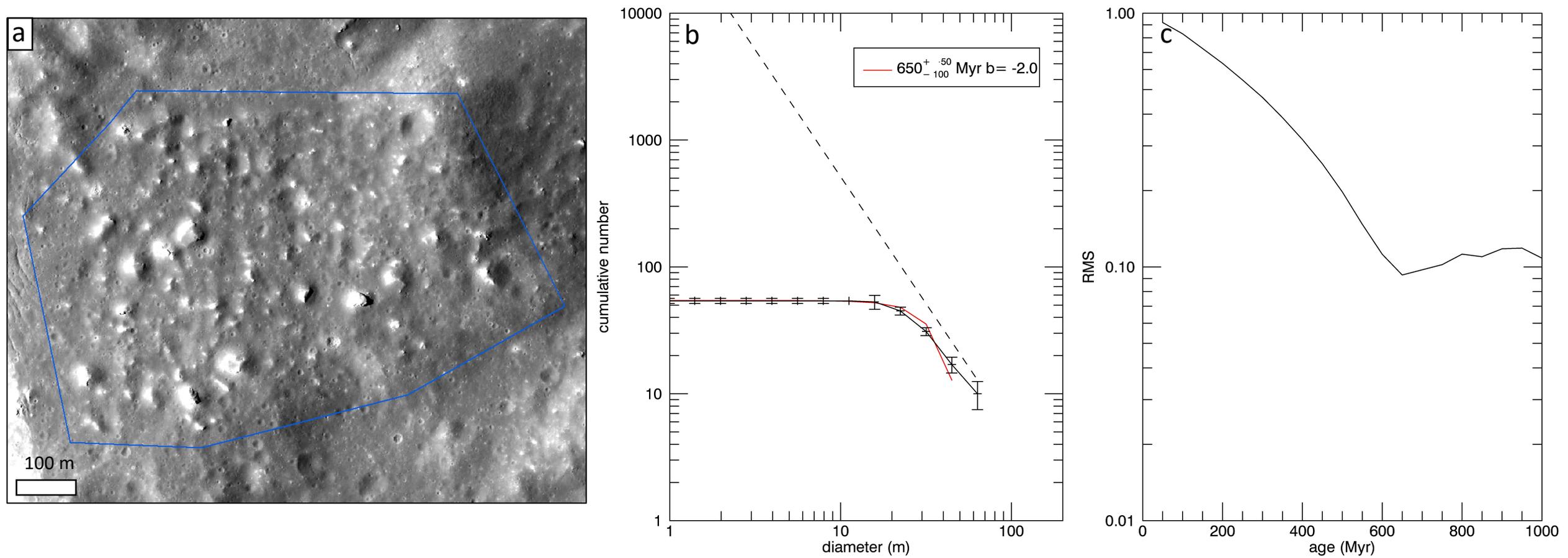

**Figure 13**. (a) NAC image M1152008432LC showing the block count area near the rim of Copernicus crater. (b) Cumulative size-frequency distribution of blocks. Data shown with crosses, model initial SFD as dashed line and best fit isochrone in red. (c) RMS deviation as a function of exposure age.

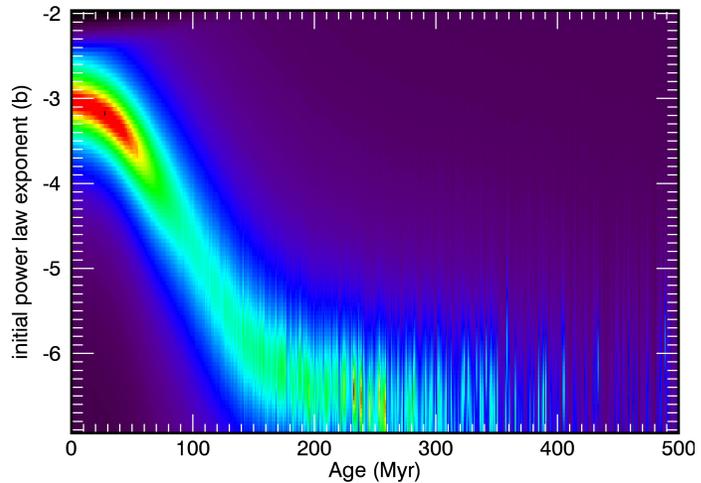 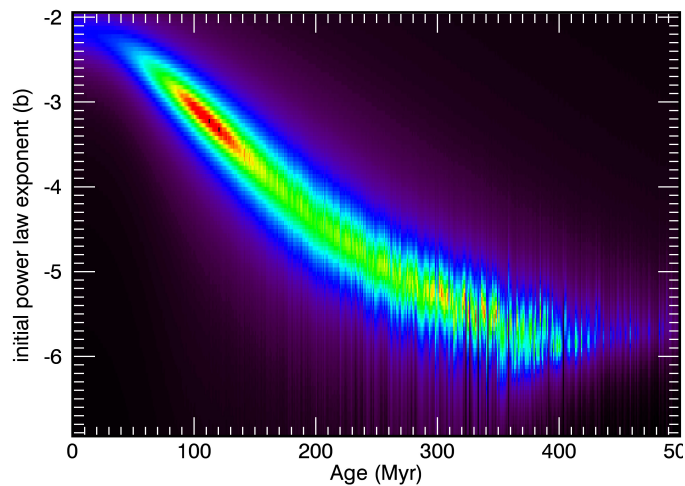 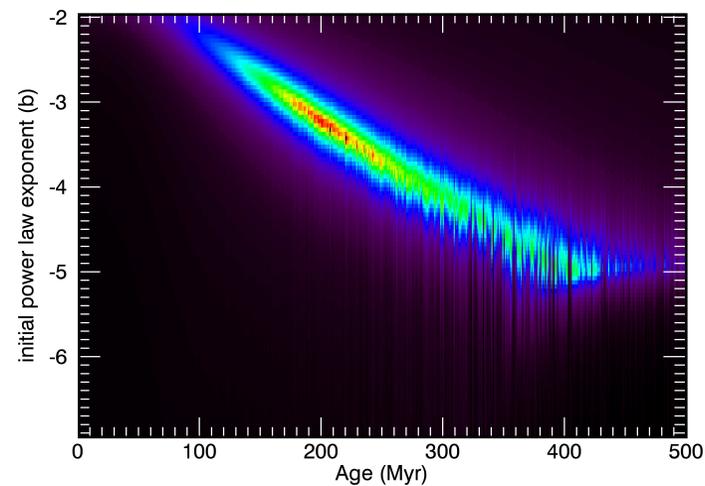 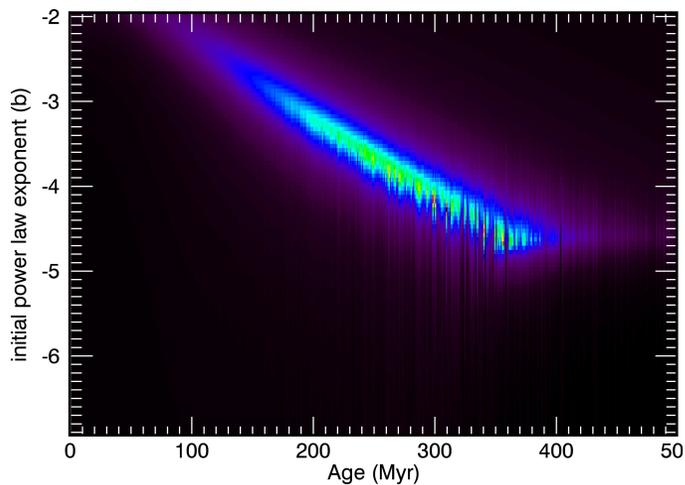 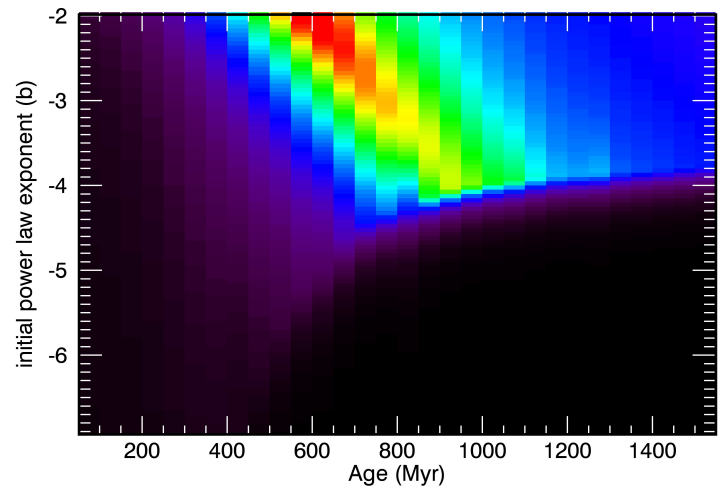 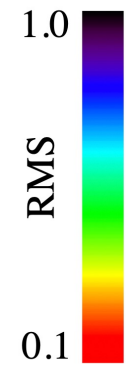

Figure 14. Color coded RMS as a function of block field age and power law exponent of the initial SFD for five lunar craters.

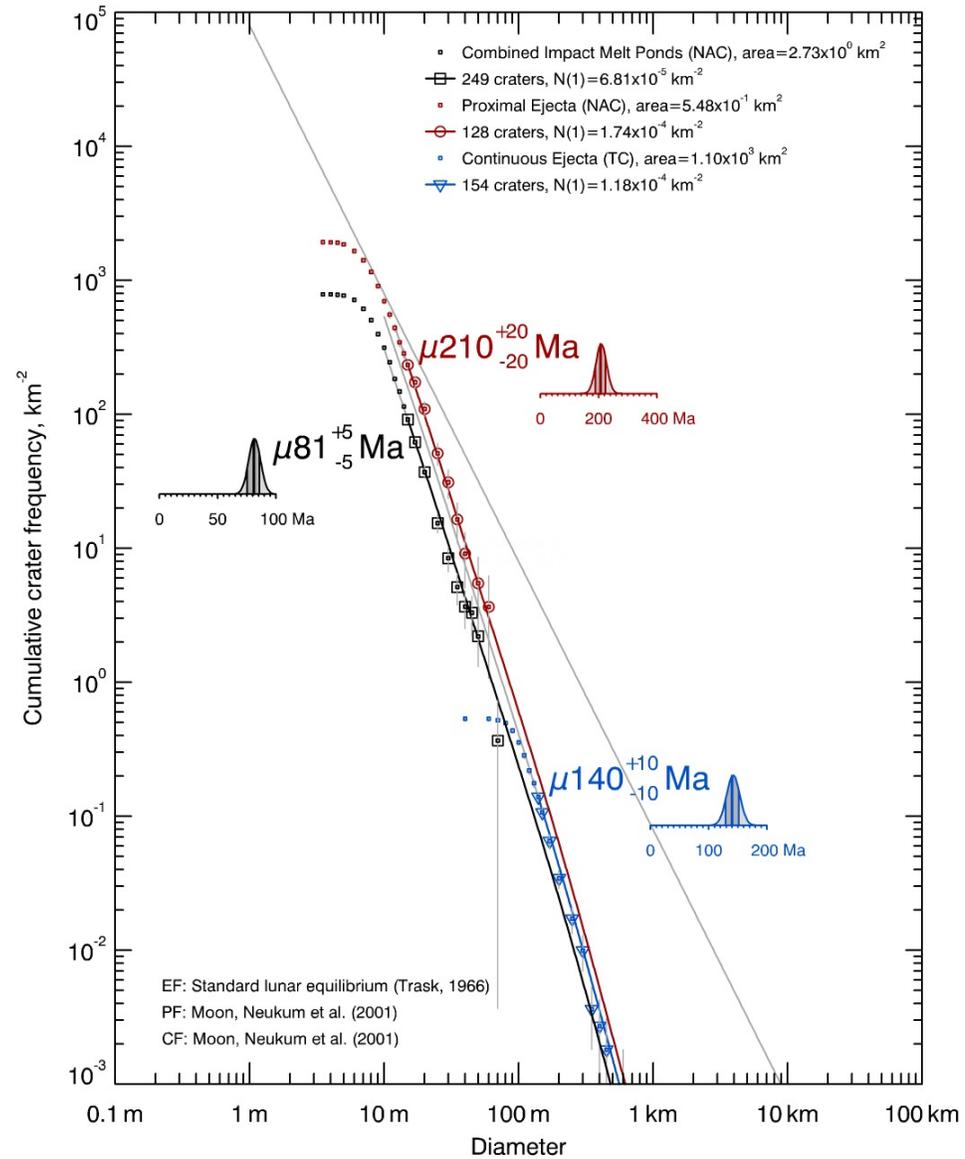

**Figure 15**. Crater SFD measurements and age estimations near Jackson crater: measurement on impact melt ponds (black), on proximal ejecta with craters <100 m in diameters (red) and distal ejecta with craters <500 m in diameter (blue). PF: Production function, CF: Chronology function.

| Crater | block SFD ages (Myr) (this study) | Range of superimposed crater SFD ages (Myr) | Reference for crater SFD ages |
|---|---|---|---|
| Byrgius A | $27^{+40}_{-27}$ | $47^{+14}_{-14}$ | Morota et al., 2009 |
| Tycho | $120^{+71}_{-51}$ | $109^{+4}_{-4}$ (radiometric age) | Drozd et al., 1977 |
| Aristarchus | $207^{+61}_{-51}$ | 164 or 280 | Zanetti et al., 2017 |
| Jackson | $298^{+1}_{-1}$ | 140–210 | this study |
| Copernicus | $650^{+50}_{-100}$ | $800^{+15}_{-15}$ (radiometric age) | Stöffler and Ryder (2001) |

**Table 1**. Overview of the ages of lunar craters estimated with the block SFD technique presented in this study, with the canonical crater SFD technique, and with the radiometric technique on samples linked to the craters. The block SFD ages have been calibrated using the radiometric age of Tycho crater.